\documentclass[12pt]{article}

\usepackage{graphicx}

\title{The scalar exotic resonances X(3915), X(3960), X(4140)}
\author{A.M.Badalian and  Yu.A.Simonov \\
NRC ``Kurchatov Institute'' \\
Moscow, Russia}

\newcommand{\beq}{\begin{eqnarray}}
 \newcommand{\eeq}{\end{eqnarray}}
\newcommand{\be}{\begin{equation}}
 \newcommand{\ee}{\end{equation}}

 \def\la{\mathrel{\mathpalette\fun <}}

\def\fun#1#2{\lower3.6pt\vbox{\baselineskip0pt\lineskip.9pt
\ialign{$\mathsurround=0pt#1\hfil ##\hfil$\crcr#2\crcr\sim\crcr}}}

\newcommand{{\SD}}{\rm SD}

\newcommand{{\Mc}}{\mathcal{M}}

\newcommand{\vep}{\mbox{\boldmath${\rm p}$}}

\begin{document}
\maketitle
\begin{abstract}
The scalar resonances $X(3915), X(3960), X(4140)$ are considered as exotic four-quark states: $cq\bar c \bar q, cs\bar c \bar s, cs\bar c\bar s$,  while
the $X(3863)$ is proved to be the $c\bar c, 2\,^3P_0$ state. The  masses and the widths of these resonances are calculated
in the framework of the Extended Recoupling Model, where a four-quark system is formed inside the bag and has relatively small size ($\la 1.0$~fm). Then the resonance $X(3915)$ appears due to the transitions: $J/\psi\omega$ into $D^{*+}D^{*-}$ (or $D^{*0}\bar D^{*0})$  and back, while the $X(3960)$ is created due to the transitions  $D_s^+D_s^-$  into $J/\psi\phi$ and back, and the $X_0(4140)$  is formed in  the transitions $J/\psi\phi$ into $D_s^{*+}D_s^{*-}$ and back. The characteristic feature of the recoupling mechanism is that this type of  resonances can be predominantly in the $S$-wave decay channels and has $J^P=0^+$. In two-channel case  the resonance occurs to be just near the lower threshold, while due to coupling to third channel (like the $c\bar c$ channel) it is
shifted up and lies by (20--30)~MeV above  the lower threshold. The following masses and widths are calculated:
$M(X(3915))=3920$~MeV, $\Gamma(X(3915))=20$~MeV;  $M(X(3960))=3970$~MeV, $\Gamma(X(3960)=45(5)$~MeV,  $M(X_0(4140))= 4120(20)$~MeV, $\Gamma(X_0(4140))=100$~MeV,  which are in good agreement with experiment.

\end{abstract}

\section{Introduction}

In the region (3.9--4.2) GeV  there are now three scalar resonances and the $X(3915)$ was the first, observed by the Belle in the $e^+e^-\rightarrow J/\psi \omega K$ process \cite{1}. Later this resonance was confirmed by the BaBar \cite{2} and in several other experiments \cite{3}), in particular, in two-photon collisions \cite{4,5}. For some years this resonance was assumed to be  the  conventional $c\bar c$ meson -- $\chi_{co}(2P)$, although this interpretation has called out some doubts \cite{6,7} (see discussion in the reviews \cite{8,9}) and does not agree with predictions in different relativistic potential models (RPM) \cite{10}-\cite{13}. The experimental masses  of the $X(3915)$ and $\chi_{c2}(2P)$ were found to be almost equal, while  in the RPMs
a smaller mass,  $M(2\,^3P_0)\cong  3870\pm 30$~MeV, and much larger mass difference, $\delta_{20}(2P) = M(\chi_{c2}(2P) - M(\chi_{c0}(2P)\cong (70- 100)$~MeV, were predicted.  Notice that  large mass difference $\delta_{20}$ is kept even if the coupling of the $\chi_{c0}(2P)$  to open channels is taken into account \cite{14,15}. Such theoretical expectations were supported by the Belle  observation of the wide  scalar $X(3860)$ resonance \cite{16}, both in $e^+e^-\rightarrow  J/\psi D^+D^-$ and $e^+e^-\rightarrow J/\psi D^0\bar D^0$  decays, which has the mass $M= 3862 ^{+26}_{-32} ~^{+40}_{-82}$
MeV and large width $\Gamma\cong 200$~MeV. The existence of the scalar $X(3860)$ resonance is confirmed by the analysis of  two-photon production,  $\gamma\gamma\rightarrow D\bar D$ in  \cite{17}.

Very recently the LHCb  \cite{18} has observed two more scalar resonances  $X(3960), X_0(4140)$ in the $D_s^+D_s^-$ mass spectrum in the $B^+\rightarrow D_s^+D_s^-K^+$ decays with the  parameters:  $M(X(3960))= (3956\pm 5\pm 10)$~MeV, $\Gamma(X(3960))= (43\pm 13\pm 8)$~MeV, $M(X_0(4140))=(4133\pm 6\pm 6)$~MeV, $\Gamma(X_0(4140))=(67\pm 17\pm 7)$~MeV, both with $J^{PC}=0^{++}$. These new scalar
resonances evidently look as exotic states and  the $X(3960)$  was interpreted as the molecular $D_s^+D_s^-$ state within the QCD sum rules approach \cite{19,20} and in a  coupled-channel
model \cite{21}; in \cite{22} it appears due to the triangle singularity, while in \cite{23} the parameters of the $X(3960)$, as a diquark-antidiquark state, were obtained in a good agreement with experiment, using the QCD sum rules approach. Notice that the masses of the $X(3960)$ and $X(4140)$  resonances lie by $\sim 20$ MeV above the thresholds:  $D_s^+D_s^-$ and $J/\psi\phi$, respectively.

In our paper we assume that the $ X(3915)$ and both the $X(3960), X_0(4140)$ belong to exotic four-quark states  $cq\bar c\bar q$ and  $cs\bar c\bar s$ and to define their parameters  we will use the Extended Recoupling Model (ERM), recently suggested in \cite{24}, which develops the  Recouplimg Model,  presented earlier \cite{25}. The ERM allows to calculate  the mass and width of a scalar four-quark states, however, within suggested mechanism such resonances cannot exist in the systems with two identical mesons, like $D_s^+D_s^+, D_s^{*+},D_s^{*+}$. This theoretical prediction is supported by the Belle experiment \cite{26}. In the ERM the system of two mesons, e.g. ($J/\psi+\phi$), can transfer into another pair of the mesons ($D_s^+,D_s^-$) by rearranging confining strings and back in the infinite chain of transformations, like $J/\psi \phi \rightarrow (D_s^+\bar D_s^-) \rightarrow J/\psi \phi \to ...$.  Note that such sequences can also be treated, for example,  in the standard OBE approximation with the meson exchanges, which, however, does not produce the singularities  near the thresholds. In the coupled-channel models (CCM) \cite{27,28} the interaction between hadrons, like $D_s^+D_s^-, J/\psi\phi$, is usually neglected, while in the ERM such interaction is taken into account, introducing the four-quark bag. It is important that all hadrons involved have rather small sizes, $\cong (0.40-0.55)$~fm and only $\omega(1S)$ has a bit larger r.m.s. $\sim 0.7$~fm. We would like to underline  the characteristic features of  the ERM \cite{24}: first, due to the string rearrangement of a four-quark system  the singularity lies close to  the lower threshold; second, this mechanism produces the resonance  in the $S$-wave hadron-hadron system and  therefore, the quantum numbers of these resonances $J^{PC}=0^{++},1^{++}, 2^{++}$; third, a resonance does not appear, if hadrons are identical.

In the literature   there are still a controversy, concerning the $X(3915)$, and different interpretations were proposed. This resonance was considered in the tetraquark model within the Born--Oppenheimer approach in \cite{29,30,31,32}, due to the triangle singularity \cite{22} and the threshold effects \cite{33}, as the molecular $D_s\bar D_s$ bound state \cite{34}, or the lightest
$cs\bar c\bar s$ state \cite{35} and as the diquark-antidiquark state, using the QCD sum rule method \cite{23,36}. In contrast to a  molecular structure of  four-quark states in the ERM these systems  are assumed  to be  compact systems, similar to the diquark-antidiquark states studied in \cite{37}. In such compact systems their wave functions at the origin are  not small and therefore they can be produced in the $\gamma\gamma$  transitions.  At this point one can assume a possible existence of at least two different but subsidiary mechanisms, producing resonances
in the four-quark and multiquark systems: first, the resonances, which are formed  inside a common multiquark bag and connected with external independent channels. As a result these resonances
could be seen in all external channels. The theory of this type of approach was suggested long ago in \cite{38}. Within the diquark-qntidiquark model the compact $Q^2\bar Q^2$ resonances were already predicted in 1988 \cite{37}.  Second type of multiquark resonances refers to the channel-coupling resonances where the internal multiquark region is only needed to connect different external channels with sufficient probability and the
considered here Extended Recoupling Model belongs to this second type.  One can easily imagine the existence of mixed type models and mechanisms where  two these dynamics interfere with each other. In what follows we shall consider only the ERM mechanism.

In our paper we will shortly discuss the higher scalars, $X(4500), X(4700)$, observed by the LHCb \cite{39}, which admit different interpretations.

The structure of the paper is as follows. In next section we shortly remind the basic formulas in two-channel case and give the values of the parameters, needed to define the masses and widths of the recoupled four-quark resonances. In section 3 more general matrix representation of the ERM is presented. In section 4 we calculate the transition amplitudes and give the masses and widths of the scalar resonances, and compare them with experimental data. In section 4  the masses of high $X(4500), X(4700)$ resonances, as the $c\bar c$ states, are discussed. Our conclusions are presented in section 5.

\section{The two-channel approach in the Extended Recoupling Model}

We study  the experimental process where, among other products, two hadrons are produced and one pair of hadrons (the pair 1) can transfer into another pair of hadrons (the pair 2). In \cite{24}
the probability amplitude of this transition was denoted as $V_{12}(\vep_1,\vep_2)$, with $\vep_1,\vep_2$ -- relative momenta of the hadrons, referring to the pair 1 and 2.
If an infinite set of the transformations was supposed and the total production amplitude $A_2$ of the pair 2 was written as a product of the slowly varying function $F(E)$ and the singular factor $f_{12}(E)= \frac{1}{1-N}$, then  the amplitude $A_2= F(E) f_{12}(E)$. This definition of the transition amplitude $V_{12}= V_{21}$ differs of that in other approaches, where   one or more the OBE diagrams with meson exchanges are taken. In the ERM \cite{24} the process occurs through the intermediate stage of the Quark Compound Bag (QCB) \cite{38,40}, where all quarks and antiquarks  of two hadrons are participating in the string recoupling and, possibly, the spin recoupling. Denoting the QCB wave functions as $\Phi(q_i)~(i=1,2,3,4)$ and the two-hadron wave functions as $\Psi_i(h_1,h_2)$, the amplitude  $V_{12}$ can be written as,
\be
V_{12}=(\Psi_1(h_{a1}h_{b1})\Phi(q_i)) (\Phi(q_i)\Psi_2(h_{a2}h_{b2})= V_1(\vep_1)V_2(\vep_2),
\label{1}
\ee
i.e.  the  amplitude  $V_{12}= \frac{1}{1- N}$  acquires the factorized form:  $V_{12}(\vep_1,\vep_2)= v_1(\vep_1) v_2(\vep_2)$ with  the factor $N$, written as

\be
N= z(E) I_1(E) I_2(E).
\label{2}
\ee
Here $z=z(E)$ can be called the transition probability, while  $I_1(E), I_2(E)$ are the following integrals (see \cite{24}):

\be
I_i(E)= v_i G_i v_i= \int{\frac{d^3 p_i}{(2\pi)^3}\frac{v_i^2(p_i)}{E'(p_i) + E^{''}(p_i) - E}}~,
\label{3}
\ee
where the hadron energies $E'(p_i),E^{''}(p_i)$ in the i-th pair near thresholds, $E'(p)=\frac{p^2}{2m'}+m'$, include corresponding thresholds $E^{th}_i$ and the reduced masses $\mu_i$, namely,
\be
E^{th}_i=m'(i)+ m^{''}(i),~~ \mu_i= \frac{m'(i)m^{''}(i)}{m'(i)+ m^{''}(i)}.
\label{4}
\ee
The result of the integration in $I_i(E)$ can be approximated by the form:
\be
I_i= {\rm const}_i \frac{1}{\nu_i-i\sqrt{2\mu_i(E-E^{th}_i)}}.
\label{5}
\ee
with $\mu_i$, defined in (\ref{4}), while  $\nu_i $ is expressed via the parameters of the hadron wave functions, which were calculated  explicitly in  \cite{24}.  Here we would like to underline  that the transition probability $z(E)$ appears to be the only fitting parameter in the ERM.

The whole series of the transitions from the pair 1 to 2 and back is summed up to the amplitude $f_{12}$,

\be
 f_{12}(E)= \frac{1}{1- z I_1 I_2}, ~~I_i= \frac{1}{\nu_i- i\sqrt{2\mu_i(E-E^{th}_i)}},
\label{6}
\ee
where $\nu_i$ are found from the four-quark wave functions, as in \cite{37,40}. The form of Eq.~(\ref{6}) takes place for the energies $E > E_1,E_2$, while
for $E < E_1,E_2$, i.e. below thresholds, the amplitude $f_1 =\left(\frac{1}{\nu_1 + \sqrt{2\mu_1(|E-E_1|)}}\right)$. It is important  that in the ERM the process proceeds with the zero relative angular momentum between two mesons, $L=0$, otherwise the transition probability $z_{12}(E)$ is much smaller and a resonance may not appear.

Note also that if the recoupling mechanism is instantaneous, or the transition from one pair of the mesons to another  proceeds instantaneously, then
the transition amplitude $V(12)$ does not factorize into $V(1)V(2)$; such an assumption was used  in the original Recoupling Model \cite{25}. However, in this approximation, e.g. for the  $T_{cc}$
resonance agreement with experiment was not reached \cite{25}. On the contrary, in the ERM \cite{24} the recoupling mechanism  proceeds in two stages: at first stage the hadrons $h_1,h_2$ collapse into
common ``compound bag"  \cite{38,40}, where the  four quarks are kept together by the confining interaction between
all possible quark pairs. This compound bag has its own wave function $\Phi_i(q_1,q_2,q_3,q_4)$ and the probability amplitude of the $h_1,h_2\to \Phi$ transition, which defines
the factor $V_1(\vep_1)$ in Eq.~(\ref{2}). In a similar way the transition from the Bag state to the final hadrons $h_3,h_4$ defines the factor $V_2(\vep_2)$ and we obtain the relation:
\be
v_1(\vep_i)= \int d^3q_1...d^3q_4 \psi_{h_1}\psi_{h_2} \Phi_i(q_1,..q_4),
\label{7}
\ee
and similar equation for $v_2(\vep_2)$, replacing $h_1,h_2$ by $h_3,h_4$. From $v_i(\vep_i)$  the function $I_i$ (\ref{3}) is defined and using (\ref{6}),  one obtains $\nu_i$.

Now we  give experimental data and corresponding the ERM parameters, referring to the  four-quark systems, $cq\bar c\bar q$ for $ X(3915)$ and $cs\bar c \bar s$ for the  $X(3960),
X(4140)$. We give also the threshold energies $E_1, E_2$. \\

{\bf {The parameters of the four-quark resonances}}

\begin{description}

\item{\bf 1)}~ $ X(3915), J^P= 0^+, \Gamma(\exp.)= 20(5)~{\rm  MeV}$ \cite{1,3}, ${J/\psi \omega  \to D^* \bar  D^*},
E_1= 3.880, E_2= 4020, \mu_1= \frac{M(J/\psi)M(\omega)}{M(J/\psi) + M(\omega)}= 0.624,~~ \mu_2 = \frac{M(D^*) M(\bar D^*)}{M(D^*) + M(\bar D^*)}= 1.050 $ (all in GeV).
From \cite{24} $\nu_1(J/\psi\omega) = 0.21$~GeV, $\nu_2(D^*\bar D^*) = 0.44$~GeV.

\item{\bf 2)}~ $ X(3960), J^P= 0^+, \Gamma(\exp.)= 43(21)~{\rm MeV}$ \cite{18},  $[J/\psi \phi] \to [D_s^-D_s^+],
E_1= 3.936,~ E_2= 4116,~ \mu_1= \frac{M_{J/\psi} M_{\phi}} {M_{J/\psi} + M_{\phi}}= 0.767, ~~\mu_2 = \frac{M(D_s^+)M(D_s^-)}{M(D_s^+ + M(D^-)} = 0.984;
\nu_1(J/\psi\phi)= 0.265,~~ \nu_2= 0.424$ (all in GeV).

\item{\bf 3)} ~ $ X(4140), J^P=0^+, \Gamma(\exp.)=67(24)~{\rm MeV} \cite{18}, ~~ [J/\psi \phi] \to [D_s^{*-}D_s^{*+}],
E_1=4.116, E_2=4.224,~ \mu_1 = 0.767,~~ \mu_2=1.056,~~ \nu_1=0.265 ,~~\nu_2 =0.410 $ (all in GeV).
\end{description}

Here $q$ can be $u,d$ quarks. To define the structure of the cross sections we start with the value of the recoupling probability  $z=0.2$ GeV$^2$ and  the parameters from the item {\bf 1)} to
obtain the distribution $|f_{12}(E)|^2$;  the values of $|f_{12}(E)|^2$ will be given  in Section 4. In the amplitude $f_{12}(E)$ the resulting singularity can be found in the form of  (\ref{6}) and for equal threshold masses it produces a pole nearby thresholds; however, real distance between the thresholds is large, $\sim 100$~MeV and the actual singularity structure can
be more complicated.

\section{The matrix approach in the ERM}

In previous Section we have presented the ERM equations in the case of two channels, which are convenient to define the mass of a resonance. However, they do not allow to study some details of the process, or to consider a larger number of channels, which can have a influence  at the properties of a four-quark system. Therefore here we present a more  general representation of the amplitude using  the unitarity relation, when the standard form of the transition amplitudes $f_{ij}(E)$ (for $L=0$) is
\be
f_{ij} - f_{ji}^* = \sum_n 2i k_n f_{in} f_{jn}^*,
\label{8}
\ee
or the unitarity relation can be realized through the $M$-matrix representation,
\be
\hat{f}_M = \frac{1}{\hat{M} - i\hat{k}},
\label{9}
\ee
where $\hat{f}, \hat{M}, \hat{k}$ are the matrices in the channel numbers \cite{28}. In some cases instead of the $\hat{M}$  it is more convenient to use the $\hat{K}$ matrix,
$\hat{M} = -\hat{K}^{-1}$, where the matrix elements (m.e.) $M_{ik}(E)$  are the real analytic functions of $E$ with the dynamical cuts. For two-channel system
$\hat{f}_M$ can be written as

\be
\hat{f}_M = \frac{1}{\hat{M} - i\hat{k}} = \frac{\hat{N}}{D(E)},
\label{10}
\ee
with
\be
\hat{N} = \left(\begin{array}{cc}
   M_{22} - ik_2 &   - M_{21}\\
      - M_{12} & M_{11} - ik_1       \end{array}\right).
\label{11}
\ee
Here
\be
D(E) = (M_{11} - ik_1)(M_{22} - ik_2) - M_{12}M_{21}.
\label{12}
\ee
One can easily establish the relation between the equations (\ref{10})- (\ref{12}) and the amplitude $f_{12}(\rm ERM)$ (\ref{6}) in two-channel case, which is a partial case of  these equations:
\be
 f_{12}(ERM) = \frac{N_{11}N_{22}}{D(E)},~~D(E) =(\nu_1 - ik_1)(\nu_2 - ik_2) - z,
\label{13}
\ee
and
\be
 z = M_{12}M_{21},~~ \nu_i\equiv M_{ii}(E).
\label{14}
\ee

One can see that for $z>0$ the values  $\nu_i=M_{ii}$ are real analytic functions of $E$. In the ERM \cite{24}  $\nu_i$ were positive constants (defined via the parameters of  the compound bag model),  while in general case Eqs.~(\ref{12})-(\ref{14}) include other transition m.e.s  $f_{ik}$. Later in our analysis we will be interested only in the denominator $D(E)$ (\ref{12}) and the factors in (\ref{13}), (\ref{14}), which fully define the position of a resonance.

The value of $z$, in principle,  can be calculated within the ERM, however, it can depend on many unknown parameters,  and at the present stage we prefer to keep $z$ as a single fitting parameter. It can be shown  that $z$ depends on the width of a resonance, but weakly depends on the resonance position.

Now we consider three channels case to study more realistic case and choose the situation, when a resonance lies  above the
threshold 3. Here we do not need to specify the channel 3, which for example,  may be a conventional $c\bar c$ state with $J^{PC}=0^{++}$. We introduce the $3\times 3$ amplitude $\hat{f}_{M}(E)$ with three thresholds $E_i~(i=1,2,3)$ and the momenta $k_i= \sqrt{2\mu_i (E - E_i)}$,
$\mu_i = \frac{m_{1i}m_{2i}}{m_{1i} + m_{2i}}$, and $E_i = m_{1i} + m_{2i}$. Here $m_{1i},m_{2i}$ are the masses of two hadrons in the channel $i$. In this case the form of Eq.~(\ref{9})
is kept,
\be
\hat{f}_3(E) = \frac{\hat{N}_3}{D_3(E)}, ~~D_3(E)=((M_{11} -ik_1)(M_{22} - ik_2) - M_{12}M_{21})) (M_{23} -ik_3) + \Delta M,
\label{eq.15}
\ee
where $\Delta M$ is
\be
\Delta M = M_{31}M_{12}M_{23} + M_{32}M_{21}M_{13} - M_{13}M_{31}(M_{22} - ik_2) - M_{32}M_{23}(M_{11}- ik_1).
\label{eq.16}
\ee
For  the energy $E$ below the thresholds, 1 and 2,   $-ik_1=|k_1|, -ik_2 = |k_2|$, and the factor $\Delta M$ is a real function of $E$. For the threshold 3  below thresholds of
1 and 2 one can define the poles of the amplitude $\hat{f}_3$, or the zeroes of $D_3(E)$, and rewrite the Eq.~(\ref{eq.15})  as,
\be
D_3 = (M_{11} - ik_1)(M_{22} - ik_2) - \tilde{z}(E),
\label{eq.17}
\ee
where the transition probability $\tilde{z}(E)$
\be
\tilde{z}(E) = M_{12} M_{21} - \frac{\Delta M (M_{33} + ik_3)}{M_{33}^2 + k_3^2}
\label{eq.18}
\ee
One can see that $\tilde{z}(E)$ acquires imaginary part, which can be of both signs. Therefore the influence of the third (or more) open
channels, lying below the thresholds $E_1,E_2$ in the $2\times 2$ matrix $f_{12}(E)$,  may be important in some cases. The channel 3 can be taken into account, introducing complex values of $z(E)$, which can depend on  the energy as in Eq.~(\ref{eq.18}).

\section{The masses and widths of the scalar resonances}

We start with the $X(3915)$ resonance and consider the following recoupling process: $J/\psi\omega \to D^* \bar D^*$. At first we look at two-channel situation and choose  the recoupling parameter $z_2=0.18$~GeV$^2$. For the $X(3915)$ structure -- $cq\bar c\bar q$ the  parameters $\mu_i, \nu_i, E_i$ are given in  the item {\bf 1}) of section 2. Then inserting all parameters to the Eq.~(\ref{13}), one obtains the distribution $|f_{12}(E)|^2~(f_2\equiv f_{12})$. Its values for different $E$ are given in Table~\ref{tab.01}, which show that the maximum takes place  at $E= 3880$ MeV, just near the lower threshold,  and  $\Gamma_2=\Gamma(2- {\rm channels})\cong 15$~MeV. In experiment for this resonance, observed by  the Belle group in the process  $e^+e^- \to e^+e^-J/\psi \omega$ \cite{1},  the larger mass $M(\exp.)=(3918.4\pm 1.9)$~MeV and $\Gamma(\exp.)=(20 \pm 5$)~ MeV \cite{3} were  obtained.

In the case of 3-channels, when  e.g. the coupling to the $c\bar c$ channel is taken into account, the factor $z_3(E)$ acquires an imaginary part. In this case we  calculate  the  amplitude $f_3(E)$, taking $z_3=(0.18 -i0.20)$~GeV$^2$;  the values of $|f_3(E)|^2 $ are given in Tab.~\ref{tab.01}.

\begin{table}[h!]
\caption{The values of the $|f_{12}(E)|^2$ for $X(3915)$ }
\begin{center}
\label{tab.01}
\begin{tabular}{|c|c|c|c|c|c|c|c|c|} \hline
$E$(GeV)        & 3.85& 3.86& 3.88 & 3.89  & 3.90 & 3.91 & 3.915 & 3.93 \\
$|f_2(E)|^2$ &3.04 & 3.68& 63.08& 25.02 & 8.33 &2.13 & 1.65 & 1.72 \\
$|f_3(E)|^2$ & 1.82 & 1.79 & 1.03 & 1.50 & 3.30  & 348.4 & 360 & 243 \\
\hline
\end{tabular}
\end{center}
\end{table}
From Table \ref{tab.01} one can see that in the 3-channel case the peak is shifted up by $\sim 35$~MeV and corresponds  the mass  $E_R\cong 3.915$ GeV and the  width  $\Gamma_3\cong  20$~ MeV,
which are in   good agreement with the experimental mass and $\Gamma(exp.)=20(5)$~MeV \cite{3}.

The scalar resonance $X(3960)$ with $J^{PC}=0^{++}$ was recently observed  by the LHCb in the  $B^+ \to  J/\psi\phi K^+$ \cite{18} and within the ERM it can be explained   due to
the infinite chain of the transitions: $J/\psi\phi \to D_s^+D_s^-$ and back. In two-channel approximation the $X(3960)$ parameters ($\nu_i, \mu_i, E_i,~(i=1,2)$ are given in the item {\bf 2)} (Section 2),  which are used to define the amplitude (\ref{13}). First, we choose  $z_2= 0.30$~GeV$^2$ and calculate the transition amplitudes  $|f_{12}(E)|^2$; their values are given in the Table ~\ref{tab.02}.

\begin{table}[h!]
\caption{The transition probability $|f_{12}|^2$ as a function of the energy $E$ for the $X(3960)$ resonance}
\begin{center}
\label{tab.02} \begin{tabular} {|c|c|c|c|c|c|c|c|c|} \hline
$E$(GeV)                  &3.85 & 3.88 & 3.89  & 3.92 & 3.95  &3.97 & 4.00 & 4.05\\

$|f_{12}|^2(z= 0.30)$    & 3.93& 28.6  &7.89   &3.20   &2.28 & 2.00  & 1.38 & 1.50\\

$|f_3|^2 (z=0.30 -i0.30)$ & 2.0 & 1.43 &  4.02  & 23.7 & 198  & 500  & 142.3  & 42.2 \\
\hline
\end{tabular}
\end{center}
\end{table}
In the two-channel approximation the numbers from  Table ~\ref{tab.02} show the  peak at $E= 3940$ MeV, near $D_s^+D_s^-$ threshold, and $\Gamma(2-ch.)\cong 15$~MeV.
In the 3-channel case  the mass of the $X(3960)$ resonance is shifted up to the position $M(3-ch.)=3970$~MeV and the width increases to the value $\Gamma(th.)\cong 45(5)$~ MeV; these values are in
agreement with the experimental numbers:  $M(X(3960))=3956(15)$~MeV,  $\Gamma(X(3960))=(43\pm 21)$~MeV \cite{18}.

In \cite{18} the LHCb has reported about another, the $X(4140)$ resonance, with $J^{PC}=0^{++}$, in the $B^+ \to D_s^+D_s^- K^+$ decay. Its  mass $M(X(4140)=4133(12)$~MeV  is close to the
$J/\psi\phi$ threshold. We consider this resonance as the $cs\bar c\bar s$ system and first calculate the squared amplitudes $|f_{12}(E)|^2$ in two-channel case,  taking the parameters $\mu_i, \nu_i, E_i$ from the item {\bf 3)} of Section 2. In this 2-channel case:  $J/\psi\phi$ and $D_s^{*+}D_s^{*-}$  the transition probability $z_2=0.35$ is taken and the calculated values of $|f_{12}|^2$  are given in Table ~\ref{tab.03}.

In three-channel case the channel  $D_s^+D_s^-$ is added as the third one, then the values $|f_3|^2$ are calculated for $z_3=0.20 - i0.20$ and given  in Table~\ref{tab.03}.

\begin{table}[h!]
\caption{The values of the $|f_{12}(E)|^2$ and $|f_3(E)|^2$ for the $X(4140)$ }
\begin{center}
\label{tab.03} \begin{tabular} {|c|c|c|c|c|c|c|}
\hline
$E$(GeV)               & 4.00& 4.07&  4.12 & 4.17& 4.22\\
$|f_{12}(E)|^2(z=0.35)$& 3.40& 8.67& 3.86  & 1.27& 0.45\\
$|f_3|^2(z=0.2-i0.2)$ & 4.54 & 12.87 & 32.12 & 13.7 & 0.66 \\
\hline
\end{tabular}
\end{center}
\end{table}
From Table~\ref{tab.03} one can see the  peak at $E_R= (4.09\pm 0.01)$~GeV, $\Gamma(th.)=60$~MeV  in two-channel  approximation  and the peak at $E_R=(4.12\pm 0.02)$~GeV with the width
$\Gamma(th.)\cong 100$~MeV in tree-channel case, which are  in good agreement with the experimental mass $M(X(4140))=(4133\pm 12)$~MeV and $\Gamma(X(4140))= (67\pm 24)$ MeV \cite{18}.

Our numbers in Tables~\ref{tab.01}--\ref{tab.03} show that in two-channel case the resonance always lies just near the lower threshold, however, if the coupling to the third channel is taken into account, then it is shifted up  and its position occurs to be close to the experimental number. The masses and widths of the exotic resonances, $X(3915), X(3960), X(4140)$, defined in the ERM,  are
given in the Table~\ref{tab.04} together with experimental data.

\begin{table}[h!]
\caption{The ERM predictions for the masses and widths (in MeV) of exotic resonances with $J^{PC}=0^{++}$ }
\begin{center}
\label{tab.04} \begin{tabular}{|c|c|c|c|c|c|}
\hline
Resonance      &$M(\rm th.)$ &  $M({\rm exp.})$  &$\Gamma({\rm th.})$ & $\Gamma(\rm exp.)$\\
\hline
$X(3915)$            &3920           & 3918 (2)     & 20  & 20(5) \cite{3}\\

$X(3960)$             &3970          & 3956(15)     & 45(5)     & 43(21) \cite{18} \\

$X(4140$)            &4120(20)      & 4133(12)     & 100   & 67(24)  \cite{18}\\
\hline
\end{tabular}
\end{center}
\end{table}
From Table~\ref{tab.04} one can see that  in the ERM the predicted masses and the widths of the scalar four-quark resonances  are  in  good  agreement with experiment, if besides
two channels, which creates the resonance, the coupling of the resonance to third channel is taken into account.

Comparing our results with those in literature, one can notice that our conclusions on the four-quark structure of the $X(3915),  X(3960, X(4140))$ also agree  with  the
analysis in the paper \cite{33}, based on the coupled channel model of the $c\bar c$ and meson-meson systems. Notice that  the general structure of the channel-coupling matrix elements in both approaches is similar.

\section{The scalar $X(4500), X(4700)$ resonances}

High scalar resonances $X(4500), X(4700)$, or $\chi_{c0}(4500), \chi_{c0}(4700)$, \cite{39}, were studied in many papers and for them two interpretations were suggested.
First, the $X(4500)$ and $X(4700)$ are considered as the $c\bar c$ states -- $4\,^3P_0$ and $5\,^3P_0$ and their masses were calculated in relativistic quark models, where coupling to open channels
was taken into account \cite{14,15,41}. In \cite{41} the influence of open channels is studied using the so-called screened potential \cite{11}, while in \cite{13} the spectrum was calculated using  the relativistic string Hamiltonian \cite{42} with the flattened confining potential \cite{43}; this flattening effect arises due to creation of virtual $q\bar q$ pairs. Notice that the flattened confining potential appears to be universal for all types of the mesons and it produces the hadronic shifts down $\sim (100-130)$~MeV for the $4P,5P$ charmonium states and gives the masses of the $4\,^3P_0, 5\,^3P_0$ states in a reasonable agreement with experiment \cite{13}. On the contrary,  in \cite{44}, within the $\,^3P_0$ model,  much smaller shifts
due to the coupled-channel effects, $\la 30$~MeV , were obtained for  the $4\,^3P_0, 5\,^3P_0$ states,  while in \cite{41}  these states acquire too large mass shifts for the chosen screened
potential.

Model-independent analysis of the $c\bar c$ spectrum  can also be done by means of the Regge trajectories, if they are defined  not for the meson mass $M(nL)$  but for the excitation energy:
$ E(nL)=M(nL)-2\bar{m}_Q$ \cite{45}, where $\bar{m}_Q$ is the current heavy quark mass \cite{13}:

\be
\left(M(n\,^3P_0) - 2 \bar{m}_c\right)^2  = 1.06 + 1.08 n_r,~(\rm in~ GeV^2); ~~n=n_r+1,~~\bar{m}_c =1.20~GeV^2.
\label{19}
\ee
This Regge trajectory gives $M(4\,^3P_0)=4.474$~GeV and $M(5\,^3P0)=4.719$~GeV, in good agreement with the LHCb data \cite{39} (see Table~\ref{tab.05}).

\begin{table}[h!]
\caption{The Regge trajectory predictions for the masses of the charmonium $n\,^3P_0$ states (in MeV) }
\begin{center}
\label{tab.05}
\begin{tabular}{|c|c|l|}
\hline
state &  $  M(nP)$   & exp. mass\\\hline

$1\,^3P_0$ &    3429& 3414.8(3)) \\
$2\,^3P_0$  & 3863    &  3862$^{+26}_{-32}$ \cite{16}\\
$3\,^3P_0$    & 4194  &  abs. \\
$4\,^3P_0$  &  4473  &  $4474\pm 6$  \cite{39}\\
$5\,^3P_0$  &   4719   &  $ 4694\pm 4 ^{+16}_{-3}$ \cite{39}\\
$6\,^3P_0$   & 4941   & abs   \\
\hline
\end{tabular}
\end{center}
\end{table}

In Table~ \ref{tab.05} the masses $ M(2\,^3P_0)=3863$~MeV, $M(4\,^3P_0)=4473$~MeV and $M(5\,^3P_0)=4719$~MeV, show very good agreement with those of $\chi_{c0}(3862)$ \cite{16},
$X(4500)$ and $X(4700)$ \cite{39}. At present other high excitations with $J^P=1^+,2^+~(n=4,5)$ are not yet found and their observation would be very important to understand
the fine-structure effects of high charmonium, in particular, the fine-structure splitting have to decrease  for a screened GE potential.

Notice that the resonance $X(4700)$ lies very close to the $\psi(2S)\phi$ threshold and this fact indicates a possible connection between the $c\bar c$ and the $cs\bar c\bar s$
states. The four-quark interpretation of the $X(4500), X(4700)$ was discussed in different models \cite{19},\cite{46}-\cite{49}, where in the mass region (4.4--4.8)~GeV
the radial or orbital excitations of a diquark-antidiquark systems can exist.

\section{Conclusions}

In our paper the scalar resonances $X(3915), X(3960), X(4140)$ are assumed to be the four-quark states, produced due to recoupling mechanism,
when one pair of mesons can transform into another pair of  mesons infinitely many times.  These resonances do not exist in the $c\bar c$ spectrum. As the four-quark states they
have several specific features:
\begin{enumerate}
\item  The resonance  appears only in the $S$-wave decay channel.
\item  Within the ERM it lies rather close to the lower threshold.
\item The scalar four-quark resonance can be created in two channel case due to transitions between channels, but it can also be coupled to another channel 3, e.g.
the $c\bar c$ channel.
\item  These resonances have no large sizes, being the compact systems, and this fact may be important for their observation. In the case of the $X(3915)$ this statement is confirmed
by the Belle analysis of the $Q^2$ distribution of the $X(3915) \to J/\psi \omega$ decays in \cite{50}.

\end{enumerate}
The masses and widths of the $X(3915), X(3960), X(4140)$, presented in Table~\ref{tab.04},  are obtained in a good agreement with experiment.

The authors are grateful to N. P. Igumnova for collaboration.

\end{document}